\documentclass[aps,pra,twocolumn,superscriptaddress]{revtex4}
\usepackage{graphicx}
\usepackage{amssymb}
\usepackage{amsmath}
\usepackage{color}
\def\cp#1{\mathbf{#1}}
\begin{document}

\title{Polarons in Ultracold Fermi Superfluids}
\author{Wei Yi}
\email{wyiz@ustc.edu.cn}
\affiliation{Key Laboratory of Quantum Information, University of Science and Technology of China,
Chinese Academy of Sciences, Hefei, Anhui, 230026, People's Republic of China}
\affiliation{Synergetic Innovation Center of Quantum Information and Quantum Physics, University of Science and Technology of China, Hefei, Anhui 230026, China}
\author{Xiaoling Cui}
\email{xlcui@iphy.ac.cn} \affiliation{Beijing National Laboratory
for Condensed Matter Physics, Institute of Physics, Chinese Academy
of Sciences, Beijing, 100190, People's Republic of China}

\date{\today}
\begin{abstract}
We study a new type of Fermi polaron induced by an impurity interacting with an ultracold Fermi superfluid. Due to the three-component nature of the system, the polaron can become trimer-like with a non-universal energy spectrum. We identify multiple avoided crossings between impurity- and trimer-like solutions in both the attractive and the repulsive polaron spectra. In particular, the widths of avoided crossings gradually increase as the Fermi superfluid undergoes a crossover from the BCS side towards the BEC side, which suggests instabilities towards three-body losses. Such losses can be reduced for interaction potentials with small effective ranges. We also demonstrate, using the second-order perturbation theory, that the mean-field evaluation of the fermion-impurity interaction energy is inadequate even for small fermion-impurity scattering lengths, due to the essential effects of Fermi superfluid and short-range physics in such a system. Our results are practically useful for cold atom experiments on mixtures.
\end{abstract}

\maketitle

\section{Introduction}
Quasiparticles serve as the cornerstone of complex collective phenomena in interacting many-body systems. As an example, the polaron, an impurity dressed by particle-hole excitations from the environment, is a typical quasiparticle that has stimulated much interest in the study of solid-state and cold-atom systems with highly polarized components. In cold atoms, polarons have been successfully explored in both the attractive and the repulsive branches of impurities interacting with a Fermi sea of identical fermions~\cite{Zwierlein, Salomon, Grimm, Kohl}. These studies are crucial for understanding the nature of normal states either against pairing physics in the case of strong attraction~\cite{Chevy, Lobo, Combescot1, Combescot2, Prokofev, Punk}, or in the context of itinerant ferromagnetism with strong repulsion~\cite{Cui, Troyer, Bruun}. However, much less is known about the fate of impurities immersed in a Fermi superfluid, where the impurity can be dressed by pairs of particles in the superfluid~\cite{Nishida}. Given the growing number of experiments on atomic mixtures, it becomes pressing to understand the fundamental physics of the underlying quasiparticles, such as polarons, in these systems.

Particularly, in the recent ENS experiment, a mixture of Bose and Fermi superfluids has been realized~\cite{Salomon2}. The two superfluids not only coexist, but also interact with each other, constituting the most charming character of this system. How to characterize the interaction energy between these two superfluids, however, is still an open question. Exact three-body calculations show that, in the BEC limit of fermions, the interaction between a bosonic atom and a molecule of two fermions cannot be faithfully described by the mean-field theory, even when the boson-fermion scattering length is small~\cite{Cui2,ZZZZ}. It then becomes essential to investigate the validity of the mean-field theory on the many-body level throughout the whole BCS-BEC crossover regime of fermions, and identify the underlying mechanism wherever it fails to apply. Here, we approach this goal via the study of polarons.

We use the variational approach to investigate the polaron physics when an impurity is immersed in a Fermi superfluid with tunable interactions. With a pairing superfluid as the background, the polaron wave function naturally acquires trimer-like terms which originate from the impurity-induced pair-breaking processes in the superfluid bath. As a result, the polaron spectrum is composed of both impurity- and trimer-dominated solutions. Due to the couplings inbetween, many avoided level crossings occur, whose widths depend sensitively on both the interaction strength and the effective range of the system. In particular, as fermions are tuned from the BCS to the BEC side, the widths of the avoided crossings gradually increase, which suggests instabilities towards three-body losses. On the contrary, the widths become narrower at smaller effective ranges, which can help stabilizing the system. For small fermion-impurity scattering lengths, the second-order perturbation theory captures the essential effects of Fermi superfluid and short-range physics in the system, and consequently shows the insufficiency of mean-field descriptions. These results rectify our understanding of interaction effects in multi-component mixtures on the mean-field level, and provide guidance for maintaining stability from three-body losses in cold atom experiments on mixtures.

The paper is organized as the following: in Sec.~\ref{sec:model}, we outline the variational approach adopted in this work and provide details on the derivation of the equations used for the numerical evaluation of the polaron energy. In Sec.~\ref{sec:results}, we present the full polaron spectra and the polaron residue. We then apply the second-order perturbation theory in Sec.~\ref{sec:pert}, and discuss the insufficiency of the Hartree-type mean-field characterization of the polaron energy. In Sec.~\ref{sec:interaction} and Sec.~\ref{sec:effrange}, we study the effects of fermion-fermion interaction and the effective range on the polaron spectrum, respectively. We examine the potential polaron to molecule transition in the system in Sec.~\ref{sec:polmol}. Finally, we summarize in Sec.~\ref{sec:sum}.

\section{Model}\label{sec:model}

We start from the Hamiltonian of our system $\mathcal{K}=H-\mu N_F$ ($N_F$ is the total number of fermions):
\begin{align}
\mathcal{K}=&\sum_{\cp k,\sigma}(\epsilon_{\cp k}-\mu)a^{\dag}_{\cp k\sigma}a_{\cp k\sigma}+\frac{g_{\rm{ff}}}{V}\sum_{\cp k,\cp k',\cp q}a^{\dag}_{\cp k \uparrow}a^{\dag}_{\cp q-\cp k \downarrow}a_{\cp q-\cp k'\downarrow}a_{\cp k'\uparrow}\nonumber\\
&+\sum_{\cp k}\epsilon_{\cp k}b^{\dag}_{\cp k}b_{\cp k}+\frac{g_{\rm{fi}}}{V}\sum_{\cp k,\cp k',\cp q}\sum_{\sigma}a^{\dag}_{\cp k\sigma}b^{\dag}_{\cp q-\cp k}b_{\cp q-\cp k'}a_{\cp k'\sigma}, \label{eqn:fullH}
\end{align}
where $a_{\cp k\sigma}$ and $b_{\cp k}$ are respectively the annihilation operators for the superfluid fermions and the impurity atom with the dispersion $\epsilon_{\cp k}={\cp k}^2/(2m)$ ($\hbar$ is taken to be unity); $\mu$ is the chemical potential of the two-species ($\sigma=\uparrow,\downarrow$) spin-balanced Fermi gas. $g_{\rm{ff}}$ ($g_{\rm{fi}}$) is the bare fermion-fermion (fermion-impurity) interaction, which is related to the scattering length $a_{\rm{ff}}$ ($a_{\rm{fi}}$) via the standard renormalization relation:~\cite{pethickbook} $1/g_{\beta}=m/(4\pi a_{\beta})-1/V \sum_{\cp k} 1/(2\epsilon_{\cp k})$, where $V$ is the quantization volume and $\beta={\rm ff},{\rm fi}$. For simplicity, we only consider the case where the impurity has the same mass as that of a fermion, and interact equally with two fermion species. Our results can be straightforwardly generalized to cases with unequal masses or imbalanced interactions.

The Fermi superfluid at zero temperature can be described by the standard BCS wave function~\cite{marderbook}:
\begin{equation}
|{\rm BCS}\rangle=\prod_{\cp k} (u_{\cp k}+v_{\cp k} a^{\dag}_{\cp k \uparrow}a^{\dag}_{-\cp k \downarrow})|{\rm vac}\rangle \sim \prod_{\cp k} \alpha_{-\cp k \downarrow}\alpha_{\cp k \uparrow} |{\rm vac}\rangle.  \label{bcs}
\end{equation}
Here, $\alpha_{\cp k \sigma}=u_{\cp k} a_{\cp k \sigma}+\eta_{\sigma}  v_{\cp k}a^{\dag}_{\cp{-k} \bar{\sigma}}$ is the annihilation operator for the Bogoliubov quasiparticles, where $\eta_{\downarrow}=-\eta_{\uparrow}=1$, $u_{\cp k}=\sqrt{(E_{\cp k}+\epsilon_{\cp k}-\mu)/(2E_{\cp k})}$, and $v_{\cp k}=\sqrt{1-u_{\cp k}^2}$; the corresponding quasiparticle energy $E_{\cp k}=\sqrt{(\epsilon_{\cp k}-\mu)^2+\Delta^2}$ with $\Delta$ the pairing order parameter.

To depict the impurity-induced polaron excitations in a Fermi superfluid, we adopt an ansatz inspired by that used in a highly-polarized two-component Fermi gas~\cite{Chevy,review1,review2}:
\begin{equation}
|P\rangle_{\cp Q}=\left( \psi_{\cp Q}b^{\dag}_{\cp Q}+\sum_{\cp k\cp k'}\psi_{\cp k\cp k'}b^{\dag}_{\cp Q-\cp k'-\cp k}\alpha^{\dag}_{\cp k'\uparrow}\alpha^{\dag}_{\cp k\downarrow}\right) |{\rm BCS}\rangle, \label{polaron_wf}
\end{equation}
where $\cp Q$ indicates the center-of-mass momentum of the polaron. The second term in the bracket, which effectively describes impurity-induced pair breaking in the superfluid and is therefore trimer-like, includes contributions from excitations like $a^{\dag}_{\cp k'\uparrow}a_{-\cp k\uparrow}$, $a^{\dag}_{\cp k\downarrow}a_{-\cp k'\downarrow}$, $a^{\dag}_{\cp k'\uparrow}a^{\dag}_{\cp k\downarrow}$ or $a_{-\cp k'\downarrow}a_{-\cp k\uparrow}$. As a first attempt at the problem, we only keep excitations to the lowest order~\cite{footnote_ansatz}. The existence of trimer-like terms in polaron wave functions is unique for a pairing-superfluid background, which significantly affects the polaron spectra. Trimers in a many-body background have received much attention recently in different contexts~\cite{Nishidacasmir,zfefimov,parishfflopolaron,zinnerefimov,ueda}, and ours is a new platform where such few-body effects can be observed.

The ground state solution can be obtained by minimizing $E_p=\langle P | {\cal K} | P\rangle-E_{\rm BCS}$, with $E_{\rm BCS}=\sum_{\cp k} (\epsilon_{\cp k}-\mu-E_{\cp k})+\sum_{\cp k} \Delta^2/(2E_{\cp k})$. Note that it is convenient to express the fermion-fermion interaction term in ${\cal K}$ in terms of the scattering between different quasiparticle states: $\cp k,\ \cp q-\cp k \rightarrow \cp k',\ \cp q-\cp k'$. Since in the large$--k$ limit, $u_{\cp k}\rightarrow 1,\ v_{\cp k}\rightarrow 0$, only one term has finite contribution to the energy:
\begin{equation}
\frac{g_{\rm ff}}{V}\sum_{\cp q,\cp k,\cp k'}u_{\cp k}u_{\cp q-\cp k}u_{\cp q-\cp k'}u_{\cp k'} \alpha^{\dag}_{\cp k\uparrow} \alpha^{\dag}_{\cp q-\cp k\downarrow}  \alpha_{\cp q-\cp k'\downarrow} \alpha_{\cp k\uparrow}.
\end{equation}

\begin{widetext}
We then have the equations:
\begin{align}
&(E_p-\epsilon_{\cp Q})\psi_{\cp Q}=\frac{g_{\rm fi}}{V} \left( 2\sum_{\cp k}|v_{\cp k}|^2\psi_{\cp Q}+\sum_{\cp k\cp k'}v_{\cp k} u_{\cp k'}\psi_{\cp k\cp k'}+\sum_{\cp k\cp k'}u_{\cp k}v_{\cp k'}\psi_{\cp k\cp k'} \right) \\
&A_{\cp k\cp k'}\psi_{\cp k\cp k'}=\frac{g_{\rm fi}}{V} \left(u_{\cp k'}\sum_{\cp k''}u_{\cp k''}\psi_{\cp k\cp k''}- v_{\cp k}\sum_{\cp k''}v_{\cp k''}\psi_{\cp k''\cp k'}+ u_{\cp k}\sum_{\cp k''}u_{\cp k''}\psi_{\cp k''\cp k'}-v_{\cp k'}\sum_{\cp k''}v_{\cp k''}\psi_{\cp k\cp k''}+v_{\cp k} u_{\cp k'}\psi_{\cp Q}+u_{\cp k}v_{\cp k'}\psi_{\cp Q} \right) \nonumber\\
&\ \ \ \ \ \ \ \ \ \ \ \ \ \ +\frac{g_{\rm ff}}{V} u_{\cp k} u_{\cp k'} \sum_{\cp k''} \psi_{\cp k'', \cp k+\cp k'-\cp k''} u_{\cp k''} u_{\cp k+\cp k'-\cp k''} .\label{eqn:poleqn2_new}
\end{align}

As $g_{\beta}$ ($\beta=\rm{ff,fi}$) would become vanishingly small after renormalization, terms including $g_{\beta}\sum_{\cp k} v_{\cp k} ...$, where $v_{\cp k}\sim 1/k^2$ at large momentum $k$, should also vanish. We therefore neglect these terms and define:
\begin{eqnarray}
A_{\cp k}&=&g_{\rm fi}(v_{\cp k} \psi_{\cp Q} +\sum_{\cp k'}u_{\cp k'}\psi_{\cp k,\cp k'});  \label{Ak_new}\\
B_{\cp k}&=&g_{\rm fi}(v_{\cp k} \psi_{\cp Q} +\sum_{\cp k'}u_{\cp k'}\psi_{\cp k',\cp k});  \label{Bk_new}\\
C_{\cp k}&=&g_{\rm ff} \sum_{\cp k'} \psi_{\cp k', \cp k-\cp k'} u_{\cp k'} u_{\cp k-\cp k'}. \label{Ck}
\end{eqnarray}
\end{widetext}

From these, we can get $\psi_Q$ and $\psi_{\cp k,\cp k'}$ in terms of $A_{\cp k}, B_{\cp k}, C_{\cp k}$, which, when plugged into Eqs.(\ref{Ak_new},\ref{Bk_new},\ref{Ck}), yield a set of closed equations:
\begin{align}
&\left(\frac{V}{g_{\rm fi}}-\sum_{\cp k'}\frac{u_{k'}^2}{A_{\cp k\cp k'}}\right)A_{\cp k}=\sum_{\cp k'} \left( \frac{2v_{k} v_{k'}A_{\cp k'}}{E_p-\epsilon_Q} +\frac{u_k u_{k'}A_{\cp k'}}{A_{\cp k\cp k'}} \right. \nonumber\\
 &\hspace{4cm}\left.+\frac{u_ku_{k'}^2C_{\cp k+\cp k'}}{A_{\cp k\cp k'}} \right); \label{eq1} \\
&\left(\frac{V}{g_{\rm ff}}-\sum_{\cp k'}\frac{u_{k'}^2u_{\cp k-\cp k'}^2}{A_{\cp k',\cp k-\cp k'}}\right)C_{\cp k}=\sum_{k'} \frac{2u_{\cp k-\cp k'}^2 u_{k'}A_{\cp k'}}{A_{\cp k',\cp k-\cp k'}},  \label{eq3}
\end{align}
with $A_{\cp k,\cp k'}=E_p-\epsilon_{\cp Q-\cp k-\cp k'}-E_{\cp k}-E_{\cp k'}$. Physically, as we only consider the case of spin-independent impurity-fermion interactions, we expect $A_{\cp k}=B_{\cp k}$, $\psi_{\cp k,\cp k'}=\psi_{\cp k',\cp k}$. We have numerically checked that this is the case.

It is also important to note that Eqs.~(\ref{eq1},\ref{eq3}) are able to reproduce the exact three-body equations when the fermion density is sent to zero. While with a finite fermion density, the interplay of Fermi superfluid and trimer physics gives rise to intriguing polaron properties as discussed below. In practice, $E_p$ and variables $A_{\cp k},C_{\cp k}$ can be numerically solved by imposing a certain momentum cutoff $k_c$. Physically, $k_c$ corresponds to setting a finite effective range $r_0\sim 1/k_c$ in the two-body collision. The coefficients in ansatz (\ref{polaron_wf}) can be easily deduced from the Schr\"odinger's equation (\ref{eqn:poleqn2_new}), which gives the polaron residue~\cite{footnote_residue}:
\begin{equation}
Z=\frac{\psi_{\cp Q}^2}{\psi_{\cp Q}^2+\sum_{\cp k,\cp k'}\psi_{\cp k\cp k'}^2}.
\end{equation}
Apparently, the polaron residue $Z\sim 1$ ($Z\sim 0$) when the system is impurity (trimer) dominated. In the rest of the work, we will focus on polarons in the ${\bf Q}=0$ sector.

\begin{figure}[t]
\includegraphics[width=9cm] {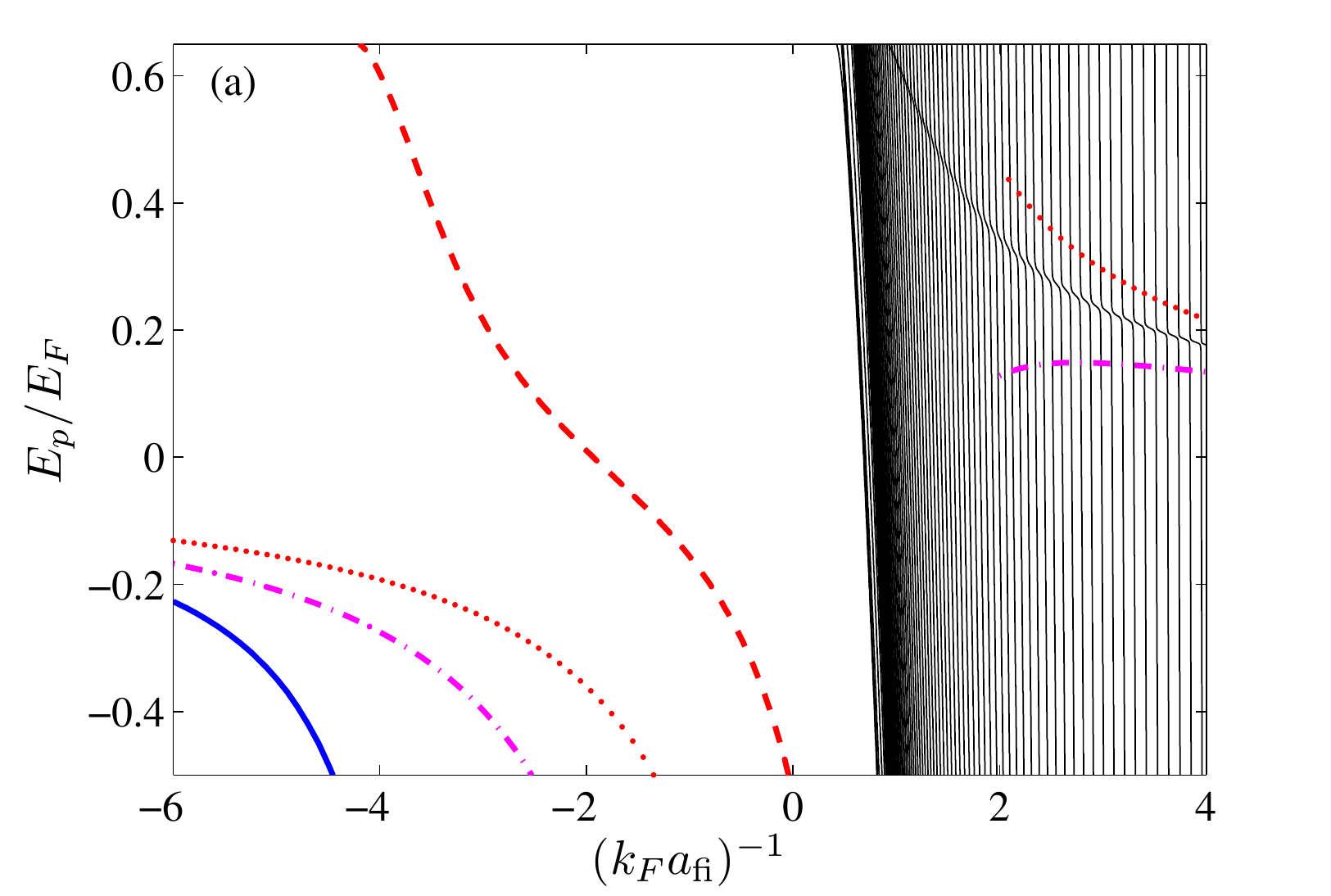}
\includegraphics[width=9cm] {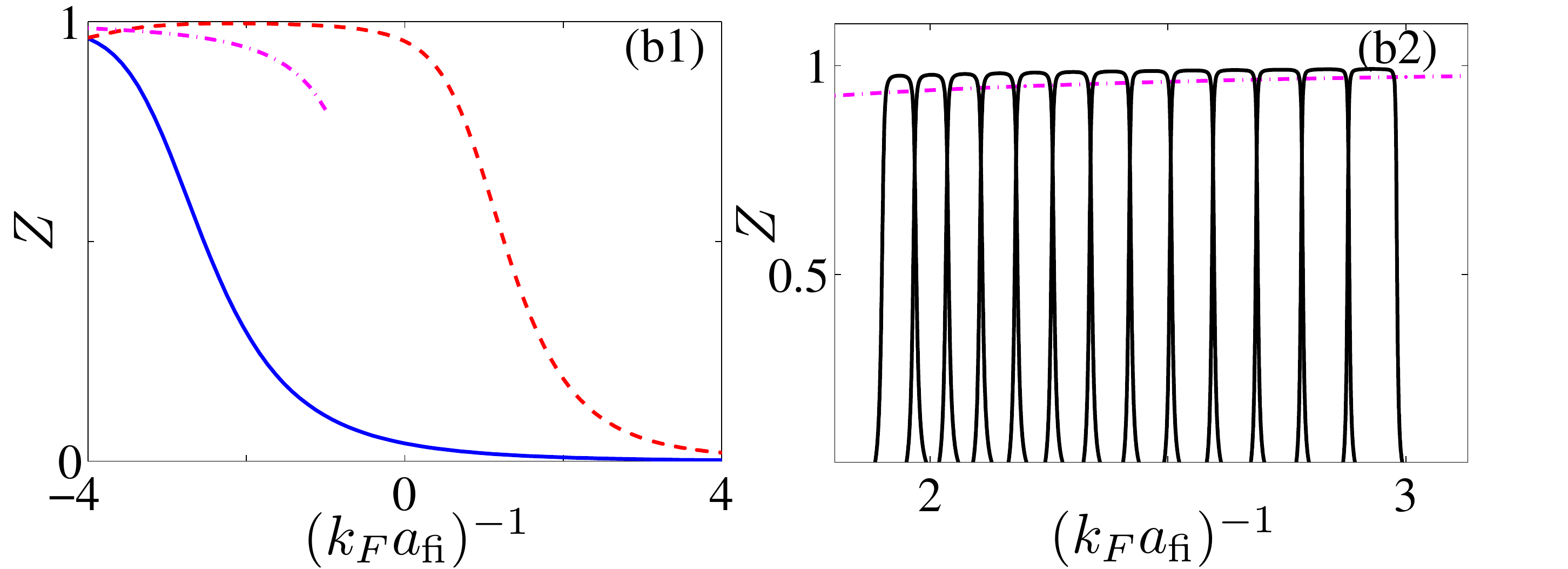}
\caption{(Color Online) Polaron spectrum (a) and residue (b1,b2) as functions of the fermion-impurity interaction strength. The fermion-fermion interaction is at resonance ($a_{\rm ff}=\infty$). (a) Blue solid, red dashed and black solid lines correspond to the lowest, second lowest and higher branches of the spectrum.
 Red dots and magenta dash-dotted line show, respectively, the perturbative energies $E_{\rm PT}/E_F$ and $(E_{\rm PT}+E'_{\rm PT})/E_F$. (b1,b2) The polaron residue for the two lowest branches (b1) and several higher branches (b2). The magenta dash-dotted curves in (b1,b2) are the residue results based on the perturbative wave function. The cutoff momentum $k_c=10k_F$, and the unit of energy $E_F=k_F^2/2m$.} \label{fig1}
\end{figure}

\section{Polaron energy and residue}\label{sec:results}
By explicitly solving Eqs. (\ref{eq1},\ref{eq3}), we get the typical polaron spectrum for a unitary Fermi superfluid (see Fig.~\ref{fig1}(a)). The spectrum exhibits many avoided level crossings between impurity- and trimer-dominated solutions, which are also apparent in the residue plots in Fig.~\ref{fig1}(b1,b2). Particularly, a wide avoided crossing exists between the two lowest branches at $(k_Fa_{\rm fi})^{-1}\sim-3$, where $k_F$ is the Fermi momentum of a non-interacting Fermi gas with the same number density as the Fermi superfluid. When $a_{\rm fi}^{-1}$ is tuned through the avoided crossing, the wave function continuously evolves from impurity dominated ($Z\sim 1$) to trimer dominated ($Z\sim 0$) in the lowest branch~\cite{footnote}. Similar atom-trimer crossover has been reported for the ground state of a two-channel model in Ref.~\cite{Nishida}, where equal fermion-fermion and fermion-impurity interactions are considered. A polaron to molecule transition~\cite{Prokofev, Punk, Nishida} can also occur within the second lowest branch of our system, when the fermion-impurity interaction is tuned towards the BEC limit.

Compared to the two lowest branches, the higher branches at positive $a_{\rm fi}$ in Fig.~\ref{fig1}(a) show more interesting properties, as multiple avoided crossings occur with widths tunable by $a_{\rm fi}$. At small $a_{\rm fi}$, the avoided crossings are very narrow, which allows us to identify a repulsive atomic branch by following the trajectory of the impurity-dominated solutions ($Z\sim 1$) while $a_{\rm fi}$ varies. The energy of the repulsive branch increases with $a_{\rm fi}$, while the width of avoided crossings becomes broader due to the enhanced coupling between the impurity and the trimer terms in Eq. (\ref{polaron_wf}). This repulsive branch eventually runs into a dense spectrum of trimer-dominated solutions.

\begin{figure}[t]
\includegraphics[width=9cm] {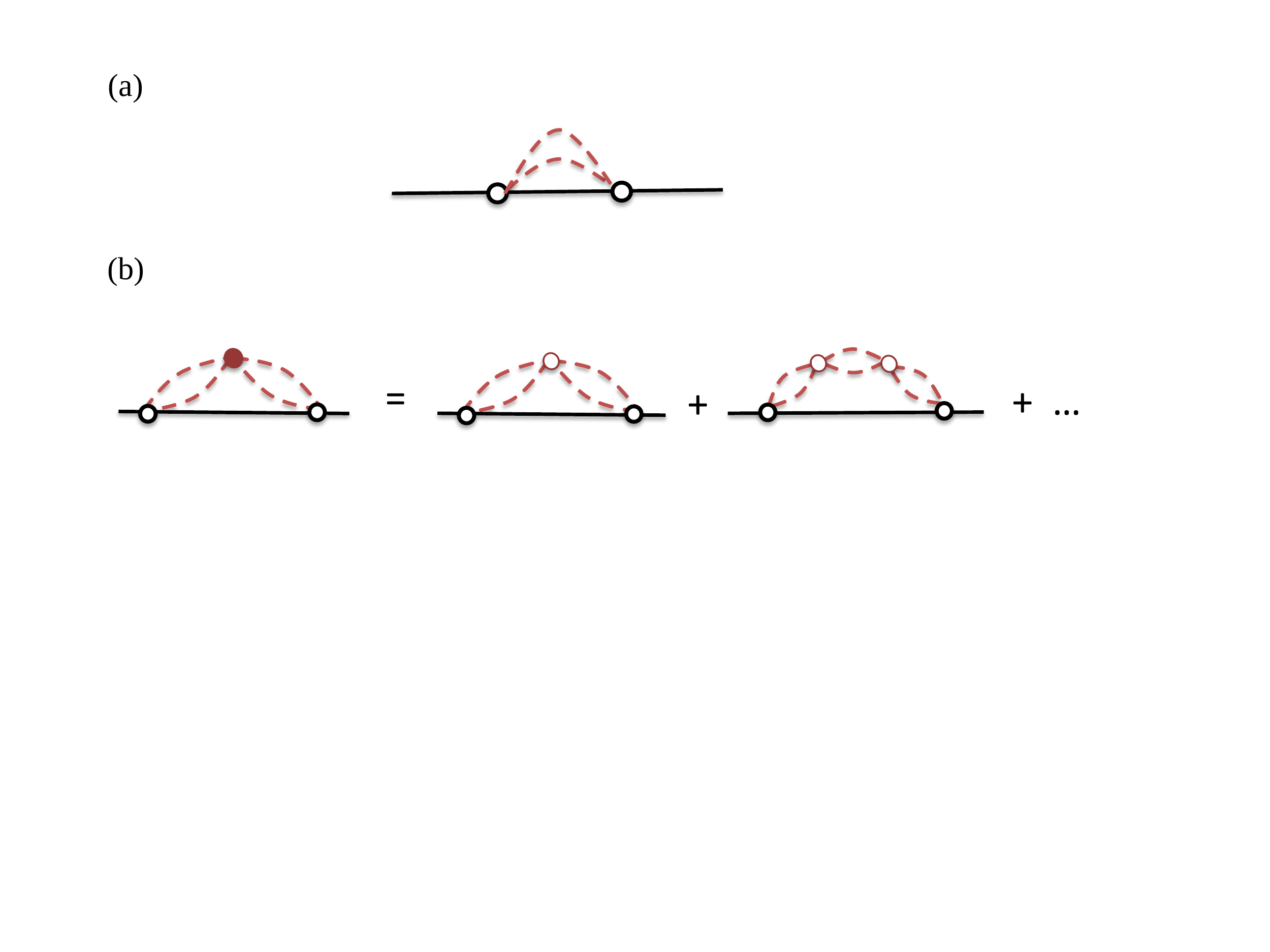}
\caption{(Color Online) Diagrams for the second-order perturbative corrections in $|a_{\rm fi}|$: (a) and (b) are, respectively, without and with the intermediate scatterings between superfluid fermions. The black solid (red dashed) line is the propagator for impurity (Bogoliubov quasiparticles). The hollow (solid) red circles represent the bare (renormalized) fermion-fermion interactions, and the  black circles represent the fermion-impurity interactions. } \label{fig_diag}
\end{figure}

\section{Perturbative Corrections}\label{sec:pert}
To gain further insights into the polaron state, we apply the second-order perturbation theory at small $|a_{\rm fi}|$. Intuitively, up to $a_{\rm fi}^2$, the perturbative energy caused by the fermion-impurity scattering can be written as:
\begin{eqnarray}
E_{\rm PT}&=& \frac{4\pi a_{\rm fi}\rho}{m} + 2(\frac{4\pi a_{\rm fi}}{mV})^2 \sum_{\cp k\cp q} \left[ v_q^2 \Big( \frac{1}{2\epsilon_k}- \right. \nonumber\\
&& \frac{u_k^2}{\epsilon_{-\cp q-\cp k}+E_q+E_k} \Big)- \left.  \frac{u_kv_ku_qv_q}{\epsilon_{-\cp q-\cp k}+E_q+E_k} \right].   \label{E_perturb}
\end{eqnarray}
Here, the first term is the mean-field contribution, with $\rho$ the total density of fermions; while the second term accounts for the intermediate scattering processes shown in Fig.~\ref{fig_diag}(a). Eq. (\ref{E_perturb}) can be continuously reduced to the case of non-interacting fermions~\cite{Cui2} with decreasing fermion-fermion interaction. However, for a unitary Fermi superfluid, $E_{\rm PT}$ is well above the polaron energy $E_p$ even at relatively small $|a_{\rm fi}|$ (see Fig.~\ref{fig1}(a)). This can be attributed to the effects of intermediate fermion-fermion scatterings, as shown diagrammatically in Fig.~\ref{fig_diag}(b). The sum of all relevant diagrams gives an additional energy correction up to the order of $a_{\rm fi}^2$:
\begin{widetext}
\begin{eqnarray}
&&E'_{\rm PT}= 2(\frac{4\pi a_{\rm fi}}{mV})^2  \sum_{k,q,k'}  \frac{(u_kv_q+u_qv_k) T_{\cp k, \cp q; \cp k', \cp{k+q-k'}} (u_{k'}v_{\cp{k+q-k'}}+v_{k'}u_{\cp{k+q-k'}})}{(\epsilon_{-\cp q-\cp k}+E_k+E_q)(\epsilon_{-\cp q-\cp k}+E_{k'}+E_{\cp{k+q-k'}})} ,\label{E_perturb2}\\
&&T_{\cp k, \cp q; \cp k', \cp{k+q-k'}}=u_{k}u_{q}u_{k'}u_{\cp{k+q-k'}} \left( \frac{mV}{4\pi a_{\rm ff}} -\sum_{\cp k} \frac{1}{2\epsilon_{\cp k}} +\sum_{\cp k''} \frac{u_{k''}^2u_{\cp{k+q-k''}}^2}{\epsilon_{-\cp q-\cp k}+E_{k''}+E_{\cp{k+q-k''}}} \right)^{-1}.  \nonumber
\end{eqnarray}
\end{widetext}
As shown in Fig.~\ref{fig1}(a), the inclusion of $E'_{\rm PT}$ considerably lowers the perturbative energy based on Eq. (\ref{E_perturb}). The remaining discrepancy is attributed to higher-order impurity-fermion scattering processes beyond the diagrams in Fig.~\ref{fig_diag}, which lead to energy corrections on the order of $a_{fi}^n$ with $n\ge 3$.

\begin{figure}[t]
\includegraphics[width=4.2cm] {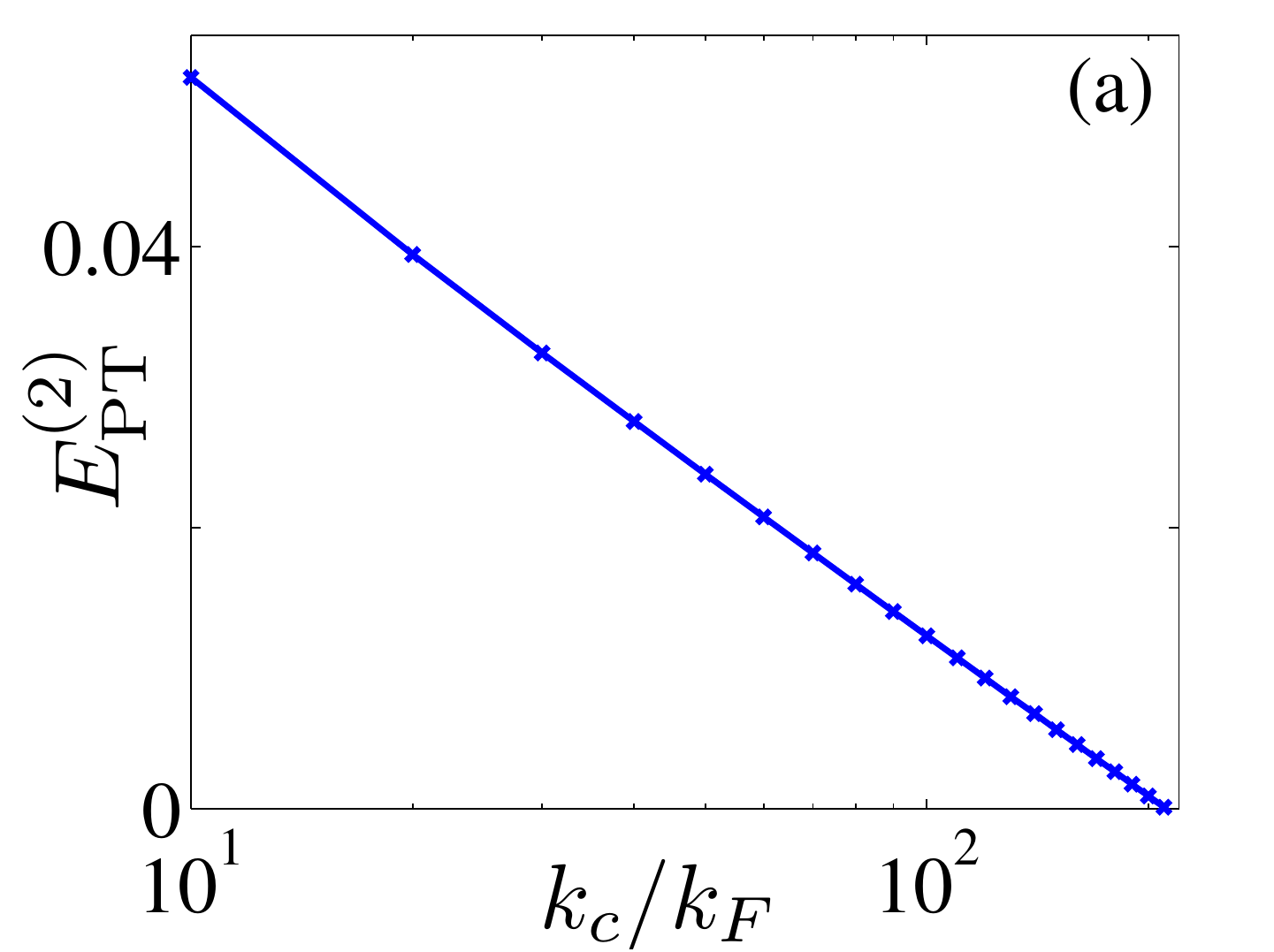}
\includegraphics[width=4.2cm] {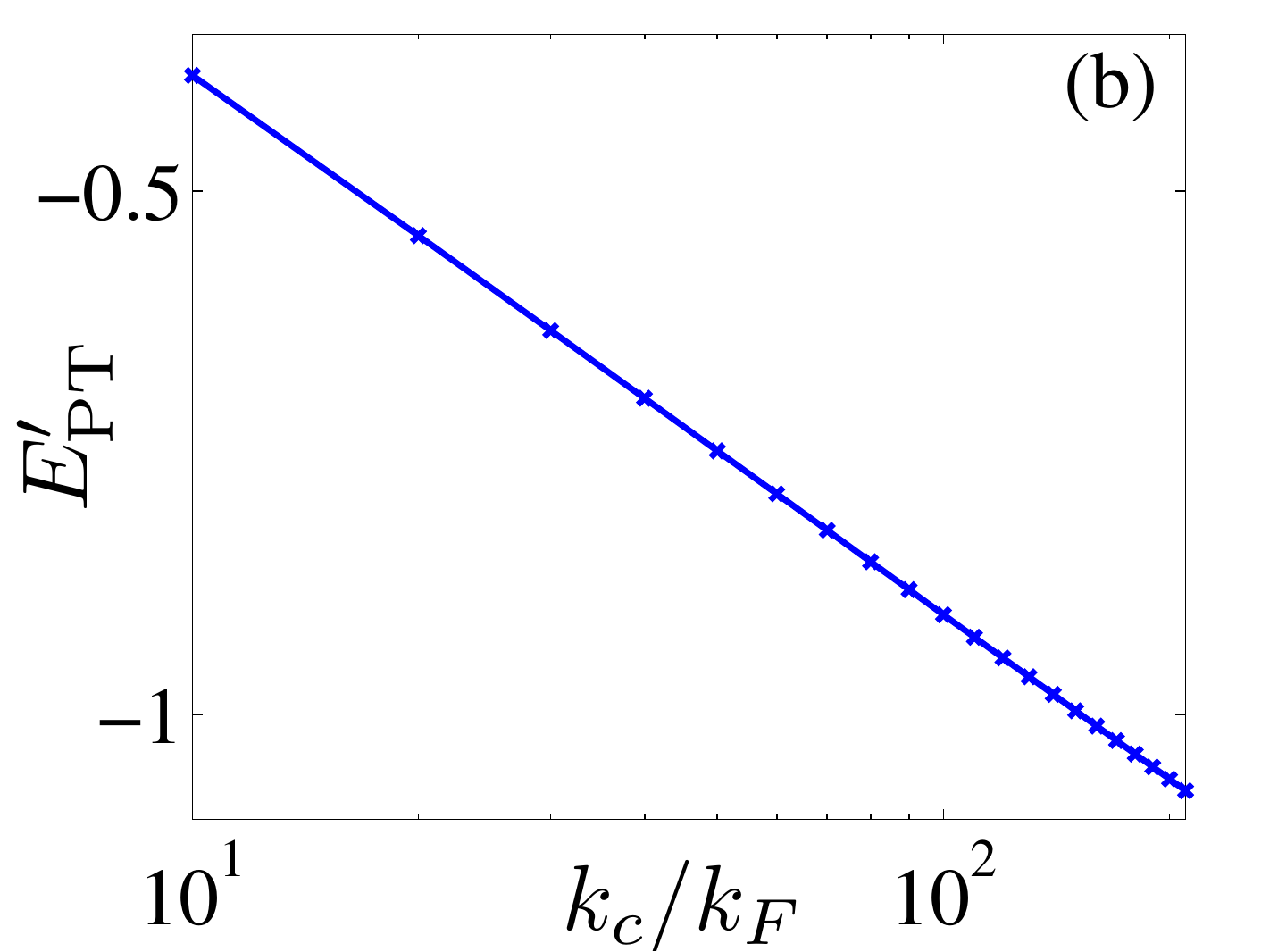}
\caption{(Color Online) The cutoff-momentum dependence of the second-order perturbative corrections: (a) is for $E_{\rm PT}^{(2)}$ (second-order terms in Eq.~(\ref{E_perturb})); (b) is for $E'_{\rm PT}$. Here we choose interaction parameters $(k_Fa_{\rm ff})^{-1}=0.1$, $(k_Fa_{\rm fi})^{-1}=2$. The unit of energy is taken to be $E_F$.
} \label{figsuppkc}
\end{figure}

The second-order perturbation results reveal several remarkable properties of polarons in the presence of pairing superfluid. First, through the inclusion of $E'_{\rm PT}$ (Eq. (\ref{E_perturb2})) and $u_k,\ v_k,\ E_k$ (Eqs. (\ref{E_perturb},\ref{E_perturb2})), the second-order terms rely crucially on the interaction between fermions. Thus, sweeping the Fermi superfluid across the resonance is expected to significantly affect the polaron. Second, the summations in $E_{\rm PT}$ and $E'_{\rm PT}$ both scale logarithmically with the momentum-cutoff $k_c$ (see Fig.~\ref{figsuppkc}). This scaling relation indicates that $E_p$ is generally non-universal, and that the mean-field evaluation of $E_p$ is inadequate even for small $|a_{\rm fi}|$. The reason of the non-universality can be attributed to the three-component nature of the system, where the high-energy (short-range) detail, or the effective range, plays an essential role. In the following, we will show the dramatic effects of the fermion-fermion interaction and the effective range on the polarons.

\begin{figure}[t]
\includegraphics[width=9cm] {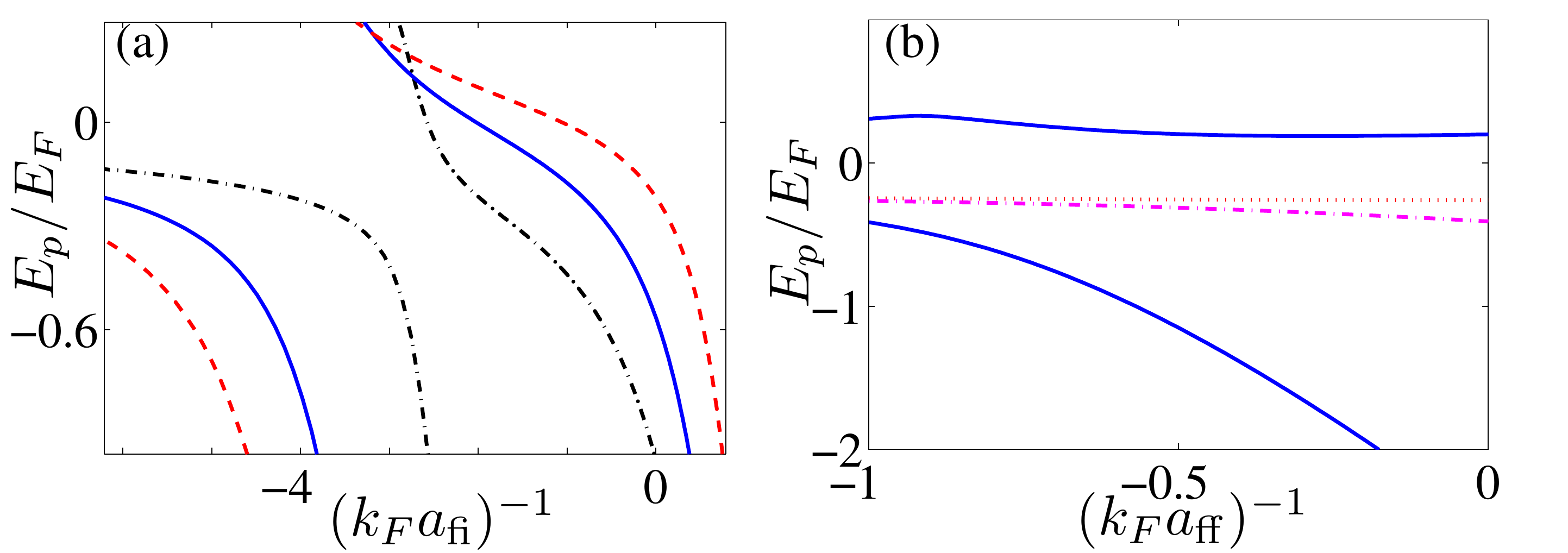}
\includegraphics[width=9cm] {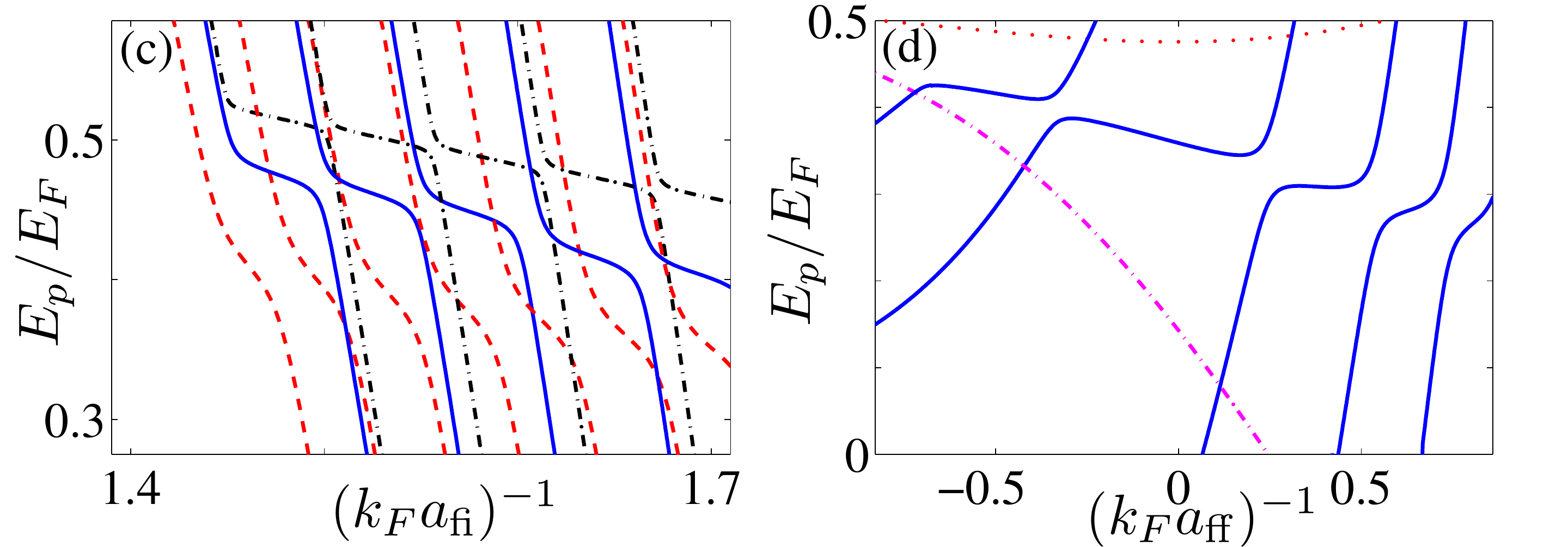}
\caption{(Color Online) Effect of fermion-fermion interaction on the polaron energy ($E_p$) for the lowest two branches (a,b) and  higher branches (c,d). (a) $E_p$ as functions of $(k_Fa_{\rm{fi}})^{-1}$, with $(k_Fa_{\rm{ff}})^{-1}=0$ (solid), $0.5$ (dashed), and $-1$ (dash-dotted). (b) $E_p$ as functions of $(k_Fa_{\rm{ff}})^{-1}$ at a fixed $(k_Fa_{\rm{fi}})^{-1}=-3$. (c) $E_p$ as functions of $(k_Fa_{\rm{fi}})^{-1}$ with  $(k_Fa_{\rm{ff}})^{-1}=0$ (solid), $0.5$ (dashed), and $-0.3$ (dash-dotted). (d) $E_p$ as functions of $(k_Fa_{\rm{ff}})^{-1}$ at a fixed $(k_Fa_{\rm{fi}})^{-1}=2$. In (b,d), the red dots and the magenta dashed-dotted line are respectively the perturbative energies $E_{\rm PT}/E_F$ and  $(E_{\rm PT}+E'_{\rm PT})/E_F$. The cutoff momentum $k_c=10k_F$.}
\label{fig3}
\end{figure}

Finally, the polaron residue $Z$ can also be estimated based on first-order (in $|a_{\rm fi}|$) wave functions derived from the perturbation theory by setting $\Psi_Q\approx 1$ (see Fig.~\ref{fig1}(b1,b2)).

\section{Effect of fermion-fermion interaction}\label{sec:interaction}
We study the variation of the polaron spectrum as the fermion-fermion interaction changes. The results are shown in Fig.~\ref{fig3}. For both the lowest two branches (Fig.~\ref{fig3}(a,b)) and the higher branches (Fig.~\ref{fig3}(c,d)), the avoided crossings become broader as fermions are tuned towards the BEC side. As a result, the repulsive atomic branch becomes difficult to identify even for small positive $a_{\rm fi}$ (Fig.~\ref{fig3}(d)), which suggests instabilities towards three-body losses.

A broad avoided crossing implies an enhanced coupling between the impurity and the trimer terms in Eq. (\ref{polaron_wf}), which we attribute to the enlarged phase space of the fermion-impurity scattering: as the fermions are more tightly paired, the Fermi surface becomes more smeared out, and the scattering phase space is less affected by the Pauli principle. Accordingly, more impurity-induced excitations emerge in the Fermi superfluid, which effectively enhances the impurity-trimer coupling in Eq. (\ref{polaron_wf}).

In addition, from Fig.~\ref{fig3}(b,d) one can see increasing deviations from the perturbative results when the fermions are tuned from the BCS to the BEC side. This suggests that a strong Fermi superfluid can dramatically affect the polaron spectra through higher-order scattering processes, which are beyond the ones shown in Fig.~\ref{fig_diag}. This is consistent with the exact calculation of atom-dimer scattering length in the deep BEC limit of fermions~\cite{Cui2, ZZZZ}, which includes all orders of scattering processes and shows the breakdown of the mean-field prediction.

\section{Effect of effective range}\label{sec:effrange}
\begin{figure}
\includegraphics[width=9cm] {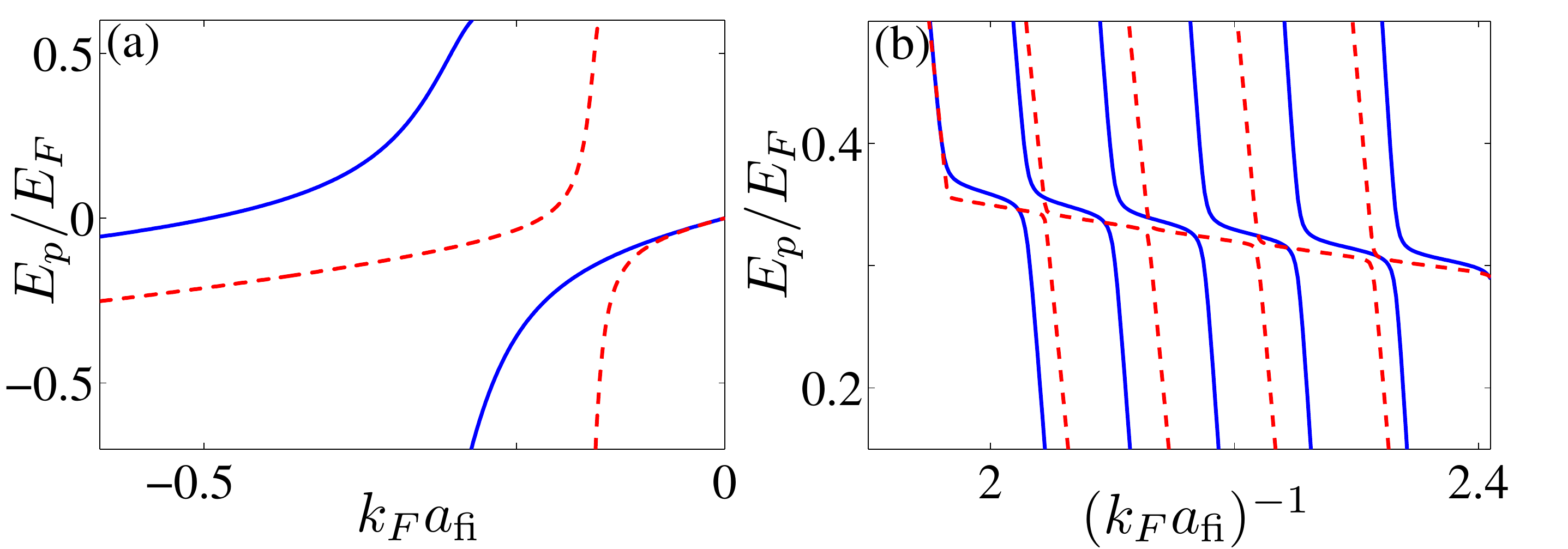}
\caption{(Color Online) Effect of effective range on the polaron spectra for the lowest two branches (a) and the higher branches (b). The fermion-fermion interaction is at resonance ($a_{\rm ff}=\infty$). The solid and dashed lines are respectively for $k_c=10k_F$ and $20k_F$. } \label{fig4}
\end{figure}

As the effective range ($r_0$) corresponds to the inverse of the cutoff momentum ($k_c$) in our formalism, its effect can be studied by changing $k_c$ in numerical simulations of Eqs.(\ref{eq1},\ref{eq3}). In Fig.~\ref{fig4}, we show the polaron spectra for two different $k_c$, where Fig.~\ref{fig4}(a) and (b) are respectively for the lower and the higher branches. A common feature is that for larger $k_c$, or smaller $r_0$, the avoided crossings move towards weaker fermion-impurity interactions, i.e., towards smaller $|a_{\rm fi}|$. Accordingly, the coupling between impurity- and trimer-dominated solutions also becomes smaller, which naturally leads to narrower avoided crossings. In practice, these results suggest that a system with a smaller $r_0$ can be more stable, as a narrower crossing with trimer-dominated solutions makes the decay into deep trimers less likely.

\section{Molecular state and polaron-molecue transition}\label{sec:polmol}

The general variational wave function for a two-body state involving the impurity atom can be written as:
\begin{align}
|M\rangle_{\cp Q}=\sum_{\cp k}\left(\varphi_{\cp k\uparrow}b^{\dag}_{\cp Q-\cp k}\alpha^{\dag}_{\cp k\uparrow}+\varphi_{\cp k\downarrow}b^{\dag}_{\cp Q-\cp k}\alpha^{\dag}_{\cp k\uparrow}\right)|{\rm BCS}\rangle.
\end{align}
In the case of a spin-independent fermion-impurity interaction, the bound state wave functions $\varphi_{\cp k\uparrow}$ and $\varphi_{\cp k\downarrow}$ are decoupled. There is then simply a two-fold degeneracy in the two-body sector. We therefore write the ansatz as:
\begin{align}
|M\rangle_{\cp Q}=\sum_{\cp k}\varphi_{\cp k\uparrow}b^{\dag}_{\cp Q-\cp k}\alpha^{\dag}_{\cp k\uparrow}|BCS\rangle.
\end{align}

From the Schr\"odinger's equation, the wave functions $\varphi_{\cp k\uparrow}$ should satisfy
\begin{align}
&E\varphi_{\cp k\uparrow}=\epsilon_{\cp Q-\cp k}\varphi_{\cp k\uparrow}+E_k\varphi_{\cp k\uparrow}\nonumber\\
&+\frac{g_{\rm fi}}{V} \left( u_{\cp k}\sum_{\cp k'}u_{\cp k'}\varphi_{\cp k'\uparrow}- v_{\cp k}\sum_{\cp k'}v_{\cp k'}\varphi_{\cp k'}+2\sum_{\cp k'}|v_{\cp k'}|^2\varphi_{\cp k} \right) ,\label{eqn:simmolorg}
\end{align}
where the last term is the Hartree term. As the Hartree terms here only serve to shift the molecular energy by a vanishingly small amount after renormalization, they are not important for the two-body bound states. Similarly, we may drop the term $g_{\rm fi}v_{\cp k}\sum_{\cp k'}v_{\cp k'}\varphi_{\cp k'}$. This leads to the equation in the two-body bound state sector:
\begin{align}
\frac{V}{g_{\rm fi}}=\sum_{\cp k}\frac{|u_{\cp k}|^2}{E_m-\epsilon_{\cp Q-\cp k}-E_{\cp k}}.\label{eqn:closedsim}
\end{align}

\begin{figure}[t]
\includegraphics[width=4.2cm] {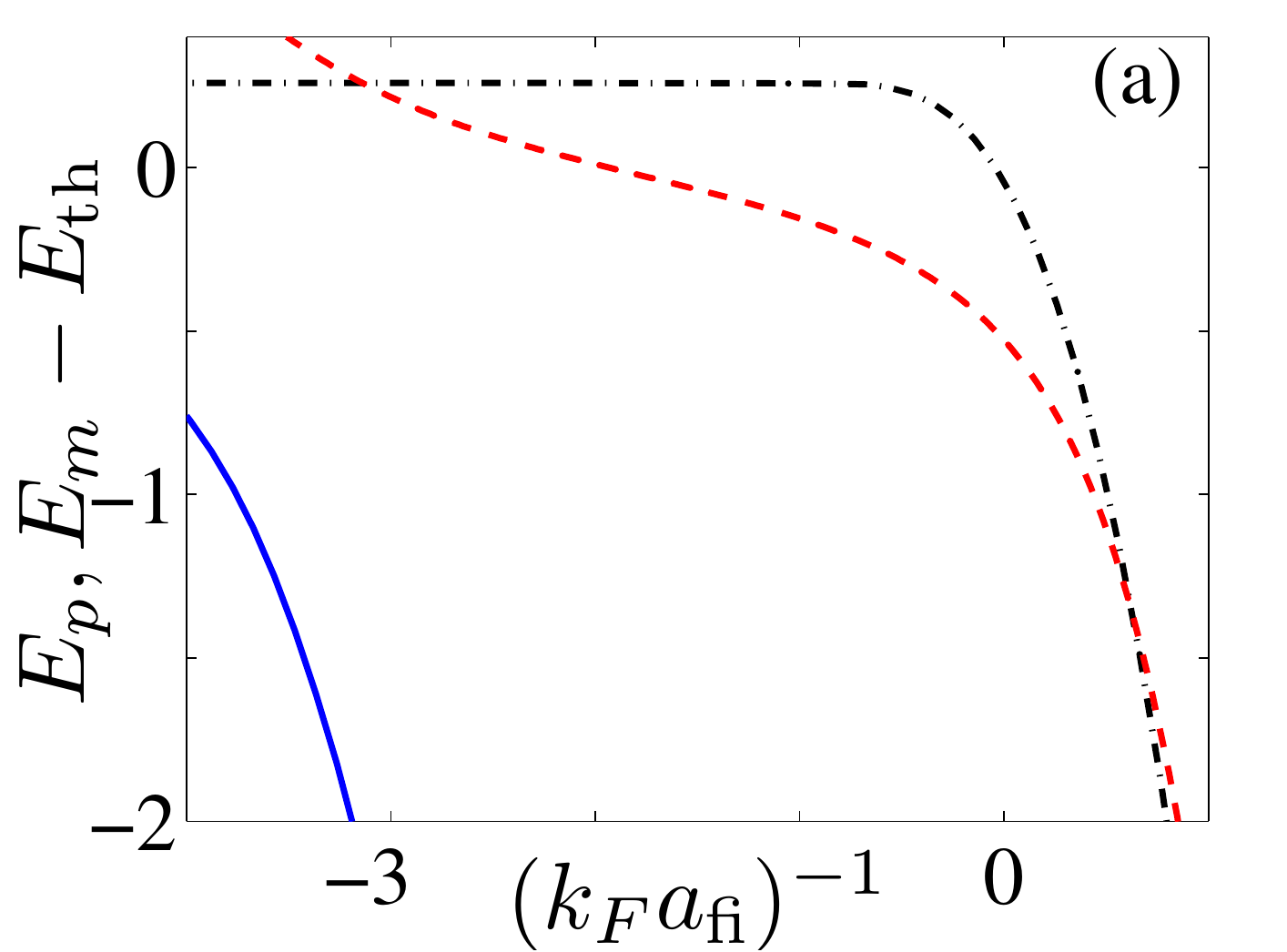}
\includegraphics[width=4.2cm] {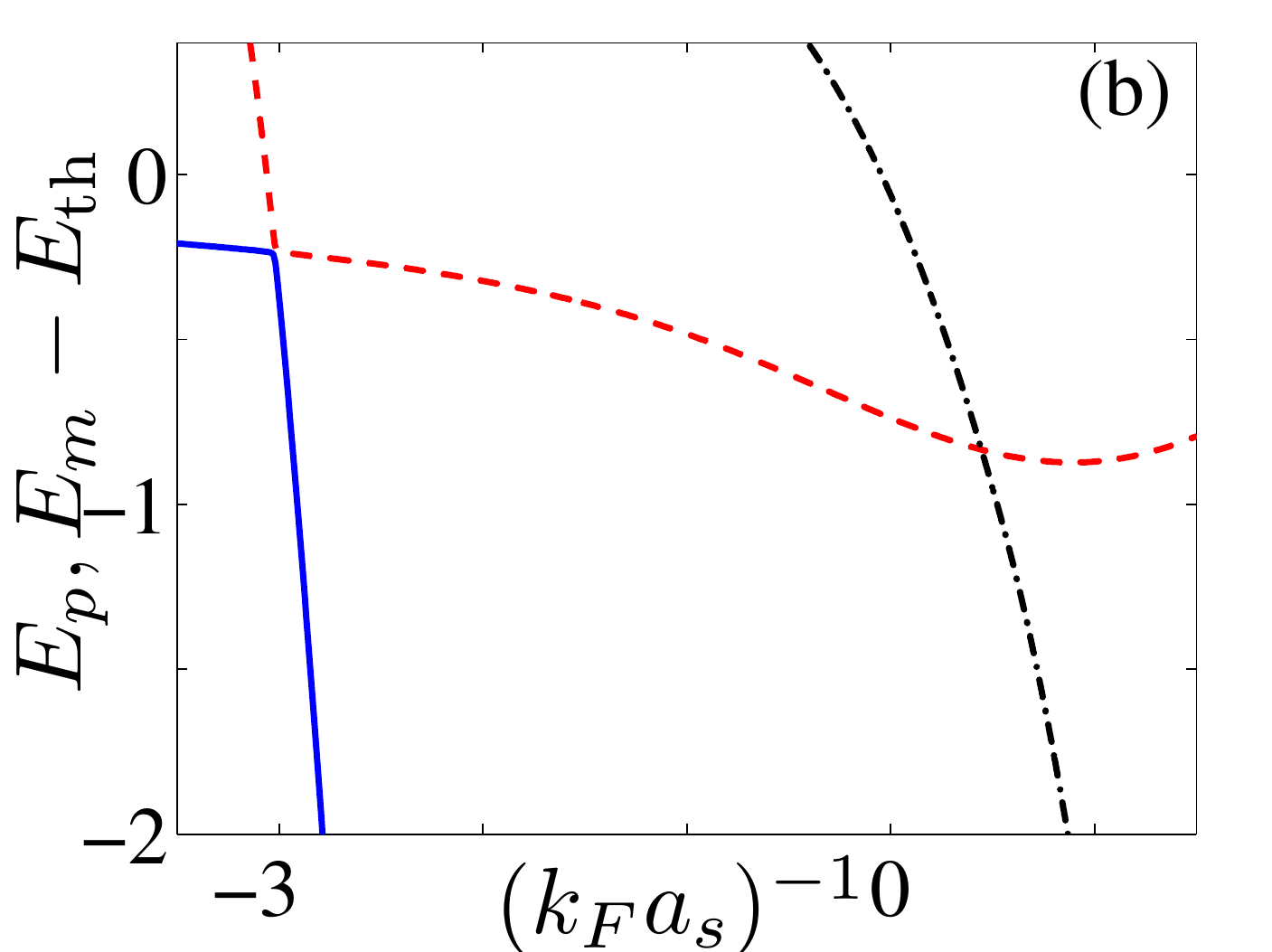}
\caption{(Color Online) (a) Polaron to molecule transition in the $Q=0$ sector. The solid and the dashed lines correspond to the lowest two branches of the polaron energy, while the dash-dotted curve is the molecular energy relative to the molecular threshold. The fermion-fermion interaction is at resonance ($a_{\rm ff}=\infty$). Cutoff momentum is chosen to be $k_c=10k_F$, and the unit of energy is $E_F$. (b) Polaron to molecule transition for the case of $a_s=a_{\rm ff}=a_{\rm fi}$, with $k_c=15k_F$.
} \label{figsuppdimer}
\end{figure}

With the zero energy reference chosen to be the BCS ground state energy, the two-body bound state threshold $E_{\rm th}$ in this sector is the quasi-particle excitation gap: $E_{\rm th}=\Delta$ when $\mu>0$ and $E_{\rm th}=\sqrt{\Delta^2+\mu^2}$ when $\mu<0$. In Fig.~\ref{figsuppdimer}(a), we show the lowest two branches of the polaron energy and the molecular energy relative to the threshold. While there are no polaron to molecule transitions in the lowest branch, which undergoes an impurity-trimer crossover before crossing with the molecular branch, there are two polaron to molecule transitions in the second-lowest branch. With increasing $k_c$, the two transition points will move towards the BCS and the BEC limit, respectively.

Similar picture for the polaron to molecule transition holds when we consider the special case where $g_{\rm ff}=g_{\rm fi}$, i.e., the fermion-fermion interaction and the fermion-impurity interaction are equal. As shown in Fig.~\ref{figsuppdimer}(b), with an appropriately chosen cutoff momentum, the avoided crossing between the lowest two branches in the polaron spectrum is very narrow. If one starts from the ground state in the BCS limit and tune the interaction strength $a_s=a_{\rm ff}=a_{\rm fi}$ fast enough, the system would not end up in the trimer-like branch beyond the avoided crossing. In this case, there will be a polaron to molecule transition on the BEC side of the resonance, at which point the ground state of the system undergoes a first-order transition from the non-universal polaron-like state into a universal molecular state. This picture is consistent with the results reported in Ref.~\cite{Nishida}.

\section{Summary and discussion}\label{sec:sum}
We have studied the polaron excitations when an impurity is interacting with a Fermi superfluid. In particular, we show the importance of higher-order scattering processes caused by the superfluid fermions in evaluating the polaron energy. Consequently, the mean-field description of polarons becomes inadequate even for small fermion-impurity scattering lengths, especially when the fermions are tuned across resonance to the BEC side. This poses new challenges on the theoretical treatment of these systems beyond the mean-field or perturbative approaches. Moreover, our work shows that the impurity-trimer coupling can be greatly enhanced by increasing the fermion-fermion and fermion-impurity interactions or by decreasing the effective range, which all lead to broader avoided level crossings. In practice, this can serve as a guideline to reduce three-body losses and maintain the stability of mixture systems. Our results can be directly probed in current cold atoms experiments. The polaron spectrum and its residue can be measured using radio-frequency spectroscopy~\cite{Zwierlein, Grimm, Kohl}, and the stability of the system can be detected through the atom-loss measurement.

Finally, we note that our polaron ansatz could be improved by including collective excitations in the Fermi superfluid due to phase fluctuations of the pairing field. This type of excitation dominates over pair breaking processes in the deep BEC regime of fermions, where the system can be modeled by a polaron on top of a molecular BEC~\cite{polaron_boson1,polaron_boson2,polaron_boson3,polaron_boson4,polaron_boson5,polaron_boson6,polaron_boson7,Demler_polaronbec,polaron_boson8}, with impurity-boson and boson-boson scattering lengths respectively given by the three-body~\cite{Cui2, ZZZZ} and the four-body~\cite{Petrov} solutions. In this work, we focus on the effects of pair breaking in the weak-coupling and the near-resonant regimes of fermions, where the collective excitations cannot qualitatively change the main features of the polaron, i.e., the emergence of trimer physics, the avoided level crossings, and the non-universality of polaron spectra.

\section*{Acknowledgements}
We thank Tin-Lun Ho, Hui Zhai, Yusuke Nishida  and Xiwen Guan for helpful feedback on the manuscript.
This work is supported by NFRP (2011CB921200, 2011CBA00200), NNSF (60921091), NSFC (11374177,11374283). XC acknowledges support from programs of Chinese Academy of Sciences. WY acknowledges support from the ``Strategic Priority Research Program(B)'' of the Chinese Academy of Sciences, Grant No. XDB01030200.


\begin{thebibliography}{99}

\bibitem{Zwierlein} A. Schirotzek, C.-H. Wu, A. Sommer, and M. W. Zwierlein, Phys. Rev. Lett. {\bf 102}, 230402 (2009).
\bibitem{Salomon}S. Nascimb\'ene, N. Navon, K. J. Jiang, L. Tarruell, M. Teichmann, J. McKeever, F. Chevy, and C. Salomon,
Phys. Rev. Lett. {\bf 103}, 170402 (2009).

\bibitem{Grimm}C. Kohstall, M. Zaccanti, M. Jag, A. Trenkwalder, P.
Massignan, G. M. Bruun, F. Schreck, R. Grimm, Nature 485, {\bf 615} (2012).

\bibitem{Kohl}M. Koschorreck, D. Pertot, E. Vogt, B. Fr\"olich, M. Feld, M. K\"ohl, Nature {\bf 485}, 619 (2012).

\bibitem{Chevy}F. Chevy, Phys. Rev. A {\bf 74}, 063628 (2006).
\bibitem{Lobo}C. Lobo, A. Recati, S. Giorgini, and S. Stringari, Phys. Rev. Lett. {\bf 97}, 200403 (2006).
\bibitem{Combescot1}R. Combescot, A. Recati, C. Lobo, and F. Chevy, Phys. Rev. Lett. {\bf 98}, 180402 (2007). 
\bibitem{Combescot2}R. Combescot and S. Giraud, Phys. Rev. Lett. {\bf 101}, 050404 (2008).  
\bibitem{Prokofev}N. V. Prokofev and B. V. Svistunov, Phys. Rev. B {\bf 77}, 125101 (2008).   
\bibitem{Punk}M. Punk, P. T. Dumitrescu, and W. Zwerger, Phys. Rev. A {\bf 80}, 053605 (2009).

\bibitem{Cui}X. Cui and H. Zhai, Phys. Rev. A {\bf 81}, 041602(R) (2010).
\bibitem{Troyer}S. Pilati, G. Bertaina, S. Giorgini, and M. Troyer, Phys. Rev. Lett. {\bf 105}, 030405 (2010).
\bibitem{Bruun}P. Massignan and G. M. Bruun, Eur. Phys. J. D {\bf 65}, 83 (2011).


\bibitem{Nishida}Y. Nishida, Phys. Rev. Lett. 114, 115302 (2015).

\bibitem{Salomon2}I. Ferrier-Barbut, M. Delehaye, S. Laurent, A. T. Grier, M. Pierce, B. S. Rem, F. Chevy, and C. Salomon, Science {\bf 345}, 1035 (2014).
\bibitem{Cui2}X. Cui, Phys. Rev. A {\bf 90}, 041603(R) (2014).
\bibitem{ZZZZ}R. Zhang, W. Zhang, H. Zhai, P. Zhang, Phys. Rev. A {\bf 90}, 063614 (2014).

\bibitem{pethickbook} C. J. Pethick and H. Smith, {\it Bose-Einstein condensation in dilute gases},  Cambridge University Press (2nd edition, 2008).
\bibitem{marderbook} M. P. Marder, {\it Condensed Matter Physics}, John Wiley \& Sons (2nd edition, 2015).
\bibitem{review1} F. Chevy and C. Mora, Rep. Prog. Phys. {\bf 73}, 11 (2010).
\bibitem{review2} P. Massignan, M. Zcaccanti, and G. M. Bruun, Rep. Prog. Phys. {\bf 77}, 034401 (2014).

\bibitem{footnote_ansatz} While this is adequate to provide quantitatively accurate results in polarized Fermi systems when compared with results from quantum Monte Carlo studies~\cite{Lobo, Combescot1, Combescot2, Prokofev, Punk}, we leave the justification of such a practice in the current system to future studies.


\bibitem{Nishidacasmir} Y. Nishida, Phys. Rev. A {\bf 79}, 013629 (2009).
\bibitem{zfefimov} D. J. MacNeill and F. Zhou, Phys. Rev. Lett. {\bf 106}, 145301 (2011).
\bibitem{parishfflopolaron} C. J. M. Mathy, M. M. Parish, and D. A. Huse, Phys. Rev. Lett. {\bf 106}, 166404 (2011).
\bibitem{zinnerefimov} N. G. Nygaard and N. T. Zinner, New J. Phys. {\bf 16}, 023026 (2014).
\bibitem{ueda} S. Endo and M. Ueda, arXiv:1309.7797.



\bibitem{footnote_residue}The residue here refers to the weight of bare impurity term in the background of the ground state Fermi superfluid. Thus, it does not include the excitation term with $\cp k+\cp k'=0$, though the impurity still stays at initial momentum $\cp Q$.

\bibitem{footnote} Close to $(k_Fa_{\rm fi})^{-1}\sim -4$, an avoided crossing exists between the second lowest and the third lowest branch. This is why the residue $Z$ in the second lowest branch does not immediately go to small value ($\sim 0$) beyond the avoided crossing with the lowest branch (see Fig.~\ref{fig1}(b1)).

\bibitem{polaron_boson1} G. E. Astrakharchik and L. P. Pitaevskii, Phys. Rev. A {\bf 70}, 013608 (2004).

\bibitem{polaron_boson2} R. M. Kalas and D. Blume, Phys. Rev. A {\bf 73}, 043608 (2006).

\bibitem{polaron_boson3} F. M. Cucchietti and E. Timmermans, Phys. Rev. Lett. {\bf 96}, 210401 (2006).

\bibitem{polaron_boson4} M. Bruderer, W. Bao, and D. Jaksch, Europhys. Lett. {\bf 82}, 30004 (2008).

\bibitem{polaron_boson5} J. Tempere, W. Casteels, M. K. Oberthaler, S. Knoop, E. Timmermans, and J. T. Devreese, Phys. Rev. B {\bf 80}, 184504 (2009).

\bibitem{polaron_boson6} S. P. Rath and R. Schmidt, Phys. Rev. A {\bf 88}, 053632 (2013).

\bibitem{polaron_boson7} W. Li and S. Das Sarma, Phys. Rev. A {\bf 90}, 013618 (2014).

\bibitem{Demler_polaronbec} F. Grusdt, Y. E. Shchadilova, A. N. Rubtsov and E. Demler, arXiv:1410.2203; Y. E. Shchadilova, F. Grusdt, A. N. Rubtsov, and E. Demler, arXiv:1410.5691.

\bibitem{polaron_boson8} R. S. Christensen, J. Levinsen, G. M. Bruun, arXiv: 1503.06979,


\bibitem{Petrov} D. S. Petrov, C. Salomon and G.V. Shlyapnikov, Phys. Rev. Lett. {\bf 93}, 090404 (2004).











\end{thebibliography}
\end{document}